\documentclass[prb,showpacs,twocolumn,preprintnumbers,amsmath,amssymb,floatfix]{revtex4}
\usepackage{graphicx}
\usepackage{color}
\usepackage{amssymb}
\usepackage{amsmath}
\usepackage{subfig}

\def\be{\begin{equation}}
\def\ee{\end{equation}}
\def\ba{\begin{align}}
\def\ea{\end{align}}
\def\bea{\begin{eqnarray}}
\def\eea{\end{eqnarray}}

\begin{document}
 \csname @twocolumnfalse\endcsname
\title{Dependence of structure factor and correlation energy on the width of
electron
wires
}
\author{ Vinod Ashokan$^1$, Renu Bala$^2$, Klaus Morawetz$^{3,4,5}$ and K. N. Pathak$^{1}$\footnote{Corresponding author Email: pathak@pu.ac.in}}
\affiliation{$^1$Centre for Advanced Study in Physics, Panjab
University, Chandigarh 160014, India \\ $^2$Department of Physics, MCM DAV College for Women, Chandigarh 160036, India \\ $^3$Muenster University of Applied Physics, Stegerwaldstrasse 39, 48565 Steinfurt, Germany\\
$^4$International Institute of Physics- UFRN,
Campus Universit\'ario Lagoa nova,
59078-970, Natal, Brazil\\
$^{5}$ Max-Planck-Institute for the Physics of Complex Systems, 01187 Dresden, Germany
} \pacs{}
\begin{abstract}
The structure factor and correlation energy of a quantum wire of thickness $b\ll a_B$ are studied in random phase approximation and for the less investigated region $r_s<1$. Using the single-loop approximation, analytical expressions of the structure factor have been obtained. The  exact expressions for the exchange energy are also derived for a cylindrical and harmonic wire. The correlation energy $\epsilon_c$  is found to be represented by $\epsilon_c (b,r_s)= \frac{\alpha(r_s)}{b} + \beta(r_s)\; ln(b) + \eta(r_s)$, for small $b$ and high densities. For a pragmatic width of the wire, the  correlation energy is in agreement with the quantum Monte Carlo simulation data.
\end{abstract}
\pacs{71.10.-w,71.10.Hf, 73.21.Hb, 71.45.Gm}

\maketitle

\section{Introduction}

The motion of electrons confined in one spatial dimension give rise to a variety of interesting phenomena with anomalous properties\cite{Giuliani05}. Recently  quasi one-dimensional systems are experimentally realized in carbon nanotubes \cite{Saito98,Bockrath99,Ishii03,Shiraishi03}, semiconducting nanowires \cite{Schafer08,Huang01} and cold atomic gases \cite{Monien98,Recati03,Moritz05}, edge states in quantum hall liquid\cite{Milliken96,Mandal01,Chang03} and conducting molecules \cite{Nitzan03}. The electrons in one dimension do not obey the conventional Fermi-liquid theory, hence the prospect of observation of non-Fermi-liquid features has given a large impetus to both theoretical and experimental research. An appropriate description  of the one-dimensional (1D) homogeneous electron gas (HEG) comes from the low-energy theory based on an exactly solvable Tomonaga-Luttinger model\cite{Tomonaga50,Luttinger63,Haldane81}. The random phase approximation (RPA) is the correct theory for HEG in the high-density limit i.e at large electron densities $n=1/(2 r_s a_B)$, with $a_B$ being the effective Bohr radius and $r_s$ is the coupling parameter.

We model the interactions by a smoothed long-range Coulomb potential $v(x) \varpropto(x^2+b^2)^{-1/2}$, where $b$ is a parameter related to the width of the wire. We also use a harmonic confinement potential. The true long-range character of the Coulomb potential has been studied by Schulz\cite{Schulz93} and Fogler\cite{Fogler05a,Fogler05} using a different approximation than RPA in certain domains of $(b,r_s)$. In fact a considerable amount of theoretical and numerical work has been done in this domain\cite{Capponi00,Fano99,Poilblanc97,Valenzuela03,Fabrizio94,Friesen80,Calmels97,Garg08,Tas03,Renu12,Renu14} using RPA and its generalized version, but still there is a need to understand the accurate parametrization of correlation energy for thin wires in the high-density limit. Therefore the calculation of the ground state energy for thin wires in the high-density limit for realistic long-range Coulomb interactions is still an open problem for 1D HEG.

Recently Lee and Drummond\cite{Lee11} studied the ground state properties of the 1D electron liquid  for an infinitely thin wire, and the harmonic wire using the quantum Monte Carlo (QMC) method, and provided a benchmark of the total energy data for a limited range of $r_s$. Furthermore, the harmonic wire with transverse confinement has been investigated with a lattice regularized diffusion Monte Carlo (LRDMC) technique by Casula et al. \cite{Casula06}, and by others\cite{Shulenburger08,Malatesta00,Malatesta99}.

Loos\cite{Loos13} has considered the high-density correlation energy for the 1D HEG using the conventional perturbation theory by taking the smoothed Coulomb potential described above in the limit $b\to 0$ (infinitely thin wire). At $b=0$ they have reported a value of correlation energy at $r_s\to 0$ of $-27.4$mHartree per electron. In their calculations the divergences in the integral for small $b$ cancels out exactly. But in RPA the divergences for $b\to 0$ and $r_s\to0$ does not cancel.

The purpose of the present paper is to study electron correlation effects in the interacting electron fluid described by RPA at high densities. The dependence of the structure factor and correlation energy on the wire-width is analyzed in the domain of $r_s<1$ and $b\ll a_B$. In this respect it is noted that RPA is a vary good approximation in the high density limit $r_s\to 0$. We have derived analytical expressions for the static structure factor in the high-density limit. The exact analytical expression for the exchange energy have also been obtained for cylindrical and harmonic potentials. It is found that on the basis of theoretical deduction and a logical assumption, the correlation energy can be represented in this region by the formula $\epsilon_c (b,r_s)= \frac{\alpha(r_s)}{b} + \beta(r_s)\; \ln(b) + \eta(r_s)$, which for small $b\to 0$ disagrees with the result obtained using conventional perturbation theory \cite{Loos13}, where it has been found that in the limit of $b\to 0$ and $r_s\to0$, the correlation energy is constant and independent of both.

The paper is organized as follows. In Section \ref{SSF1DFL}, we calculate the static structure factor within RPA and using the first-order approximation to RPA. In Section \ref{RPA_GSE}, the RPA ground-state energy formula is given. Subsection \ref{RPA_GSE} A provides the exact analytical result of exchange energy for cylindrical and harmonic potentials for finite $b$ and $r_s$. The result for small-$b$ limit is also given there. Subsection \ref{RPA_GSE} B describes the correlation energy partially by an analytical formula and partially through numerical calculation. The final result of the correlation energy and its parametrization is presented in section III. In Section \ref{Result_Discussion} we discuss the results.

\section{Structure factor}
\label{SSF1DFL}

In this section we calculated the  structure factor $S(q)$ within the RPA and its first-order version, where it is possible to obtain the result analytically. The  RPA density response function $\chi(q,\omega)$ is given by \cite{Pines66},
\be
\label{RPAresponse}
\chi(q,\omega)=
\frac{\chi_{0}(q,\omega)}{1-V(q)\;\chi_{0}(q,\omega)}
\ee
where, $V(q)$ is the Fourier
transform of the inter-electronic interaction potential. For
harmonically trapped electron wires, and for cylindrical wires it is
given\cite{Friesen80} respectively by
$V(q)=\frac{e^2}{4 \pi \epsilon_{0}}~E_{1}(b^2 q^2)~ e^{b^2 q^2}$ and
$V(q)= 2~\frac{e^2}{4 \pi \epsilon_{0}}~K_{0}(b q)$, where $E_{1}$ is
the exponential integral and $K_{0}$ the modified Bessel function of
$2^{nd}$ kind.

\begin{figure}[!t]
\begin{center}
\subfloat[{}]
{\includegraphics [scale=0.78]{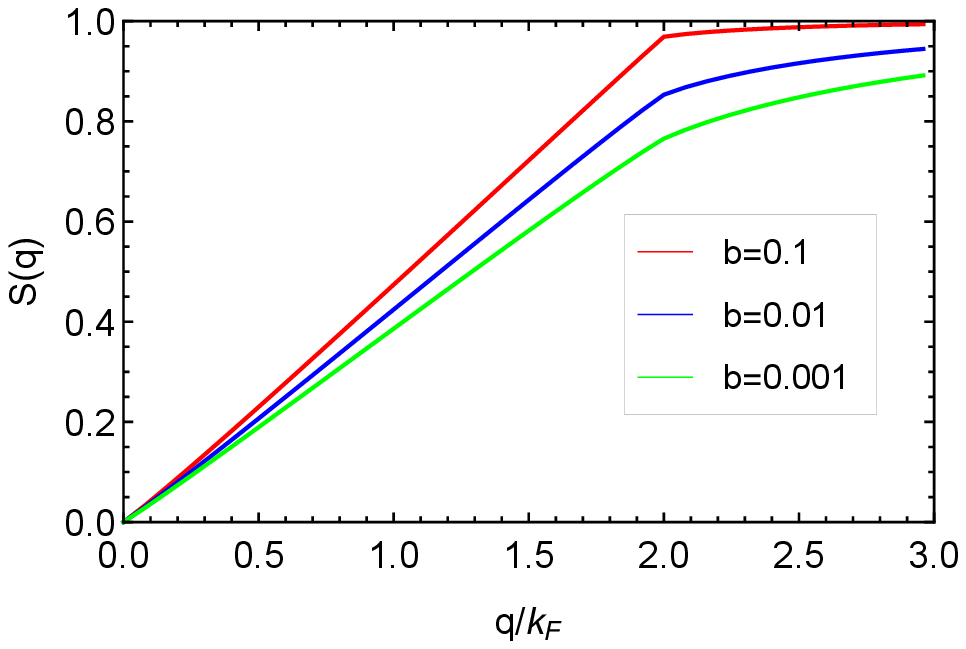}}\quad
\subfloat[{}]
{\includegraphics [scale=0.36]{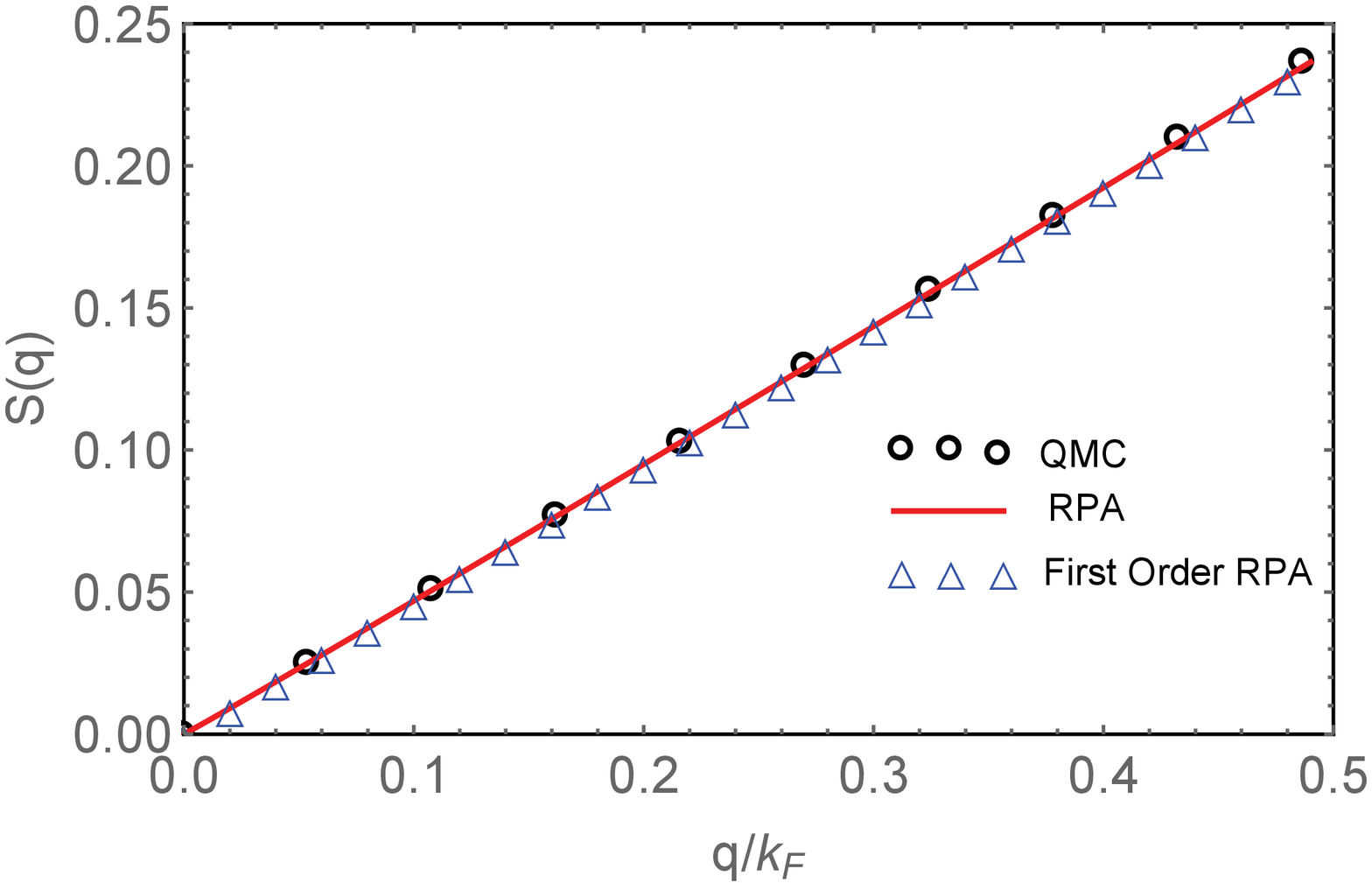}}
\end{center}
\caption{\footnotesize(Color online) $(a)$ The static structure factor $S(q)$ in RPA for a cylindrical wire is plotted as a function of $q/k_F$ for $r_s=0.3$, at different thickness of the wire $b$=0.1, 0.01 and 0.001 a.u. $(b)$ the RPA structure factor  is compared with diffusion Monte Carlo simulation, and with the  first order RPA structure factor for $r_s$=0.1 and $b$=0.025.}
\label{SqRPA}
\end{figure}

The static structure factor is defined through the fluctuation-dissipation theorem as
\be
\label{ssf}
S(q)=-\frac{1}{\pi\; n} \int_{0}^{\infty} d\omega\; \chi^{''}(q,\omega)
\ee
where $\chi^{''}(q,\omega)$ is the imaginary part of the density
response function (\ref{RPAresponse}). The integral in (\ref{ssf}) can be re-written using the
contour integration method \cite{Giuliani05} as,
\be
\label{ssfRPA}
S(q)=-\frac{1}{\pi\; n} \int_{0}^{\infty} d\omega\; \chi(q,i
\omega)
\ee
where $n=(k_F\; g_s)/\pi$ is the linear electron number density, $g_s$ is the spin degeneracy factor and $k_F$ is the Fermi wave vector. Using the high-density expansion
\be
\label{resHDE}
 \chi(q,i\omega)=\chi_{0}(q,i \omega)+\chi_{0}(q,i
\omega)\; V(q)\; \chi_{0}(q,i\omega),
\ee
where,
\bea
\label{chiomega}
\chi_{0}(q,i\omega)=\frac{g_{s} m}{2 \pi q} \ln
\bigg[\frac{\omega^2+(\frac{q^2}{2 m}-\frac{q
k_{F}}{m})^2}{\omega^2+(\frac{q^2}{2 m}+ \frac{q k_{F}}{m})^2}\bigg],
\eea
the structure factor (\ref{ssfRPA}) can be calculated for $r_s\to 0$ using (\ref{resHDE}) and (\ref{chiomega}).
The zeroth-order static structure factor is easily calculated
\bea
S_0(q)&=&-\frac{1}{n\; \pi } \int_{0}^{\infty} \chi_0(q,i\;\omega) d\omega
\nonumber\\
&=&
    \begin{cases}
    \frac{q}{2k_F},& ~ q< 2k_F \\[2ex]

    1,              & ~ q> 2k_F
\end{cases}.
\label{S0}
\eea

The first-order correction to the structure factor can be obtained by substituting $\chi_0(q,i\omega)$ in the second term of (\ref{resHDE}), and than using it in (\ref{ssfRPA}). The resulting integral can be
performed analytically and we obtain the result for $q < 2k_F$ as,
\ba
S_{1}(q)=-& v(q)\frac{g_s^2\;r_s\;2k_F}{\pi^2\; q}
\bigg[
\left (1 - {q\over 2k_F}\right )\ln\left (1-{q\over 2k_F}\right )
\nonumber\\
+&\left (1 + {q\over 2k_F}\right ) \ln{\left (1 + {q\over 2k_F}\right )}
\bigg].
\label{S1}
 \end{align}
Similarly, for $q>2k_F$ one obtains
\bea
S_{1}(q)=-& v(q)\frac{g_s^2\;r_s\;2k_F}{\pi^2\; q}
\bigg[
\left ({q\over 2k_F} -1 \right )\ln\left ({q\over 2k_F}-1\right )
\nonumber\\
+&\left (1 + {q\over 2k_F}\right ) \ln{\left (1 + {q\over 2k_F}\right )}
-{q\over k_F} \ln{q\over 2k_F}\bigg].
\label{S1a}
\eea
Here and in the following we use $V(q)=v(q) e^2/4 \pi \epsilon_0$.
In the limit of small $q$, $q$ around $2k_F$ and large $q$, the $S_1(q)$ takes the simpler forms given as
 \bea
 \label{ssflimits}
   S_{1}(q)=
    \begin{cases}
    - v(q\to0)\frac{g_s^2\;r_s}{2\pi^2}\frac{q}{k_F} ,& ~ q\ll 2k_F \\[2ex]

- v(q\to2k_F) \frac{g_s^2\;r_s}{2\pi^2}\frac{k_F}{q}\Lambda(z), & ~q \to 2k_F \\[2ex]

   - v(q\to\infty)\frac{4g_s^2\;r_s}{\pi^2}\frac{k_F^2}{q^2},              & ~ q \gg 2k_F
\end{cases}
\eea
where $\Lambda(z)=(8\ln(2) -2 |z| + 2|z|\ln|z| -
  \frac{3}{4}|z|^2)$ and  $z=\frac{q-2k_F}{k_F}$.
It can be easily seen that for harmonic wires the interaction potential approaches
\be
 v(q)=\left\{%
\begin{array}{ll}
\label{HarPoten}
   -\gamma-2\ln (b q) ~~\text{for} & \hbox{ bq $\rightarrow$ 0} \\
    1/(b q)^2 ~~~~~~~~~~ \text{for} & \hbox{ bq $\rightarrow$ $\infty$.} \\
\end{array}%
\right.\ee
where $\gamma$ is the Euler Gamma constant.

For a cylindrical potential the corresponding results are,
\be
 v(q)=\left\{%
\begin{array}{lcl}
\label{ClyPoten}
   -\gamma+\ln(2)-\ln (b q) &\text{for} & \hbox{ bq $\rightarrow$ 0} \\
   e^{-bq}\sqrt{\frac{\pi}{2bq}}&\text{for} & \hbox{ bq $\rightarrow$ $\infty$.} \\
\end{array}%
\right.\ee
Both potentials behave
similarly at the small $bq$ limit, but at large $bq$ they differ. Substituting values of $ v(q)$ from (\ref{HarPoten}) in (\ref{ssflimits}), the corresponding leading term agrees with Fogler\cite{Fogler05}.

To see the effect of thickness $b$ of the wire we calculate the structure factor from  (\ref{ssfRPA}) by using  (\ref{RPAresponse}) and (\ref{chiomega}), for $r_s<1$ and plot them in Fig.~\ref{SqRPA}. It is seen from figure \ref{SqRPA}a  that as $b$ decreases, the structure factor $S(q)$ also decreases.  A similar trend is also obtained for other $r_s$. To see the validity of the first-order structure factor we plot in Fig.~\ref{SqRPA}b for $r_s=0.1$,   $S_0(q)+S_1(q)$ and RPA for $b=0.025$. These are compared with diffusion quantum Monte Carlo simulation\cite{vinod17} for an infinitely thin wire. All three curves match perfectly. This demonstrate that the system behaves as a  gas of non-interacting electrons as conjectured by Fogler\cite{Fogler05}.

\section{Ground state energy}
\label{RPA_GSE}

The ground-state energy can be obtained by the density-density response function in conjunction with the fluctuation-dissipation theorem as \cite{Giuliani05},
\bea
\label{gsE}
E_g=E_0+\frac{n}{2}\sum_{q\neq 0}v(q) \bigg( -\frac{1}{n\pi}\int^1_0 d\lambda \int^{\infty}_0 \chi(q,i \omega;\lambda)\; d\omega\bigg).\nonumber\\
\eea
It further simplifies into a sum of kinetic energy of the non-interacting gas with the exchange energy and the residual energy (i.e correlation energy) as,
\bea
E_g&=&E_0+E_x+E_c
\eea
where
\bea
\label{exchE}E_x&=&\frac{n}{2}\sum_{q\neq 0}v(q) \bigg(-\frac{1}{n\pi}\int^1_0 d\lambda \int^{\infty}_0 \chi_0(q,i \omega)\; d\omega-1\bigg)\nonumber\\ \\
\label{corrE}E_c&=&\frac{n}{2}\sum_{q\neq0}\bigg(-\frac{1}{n\pi}\int^{1}_0 d\lambda\int^{\infty}_0 \frac{\lambda\; v(q)^2\; \chi_0^2(q,i\omega)}{1-\lambda\; v(q)\; \chi_0(q,i\omega)}d\omega\bigg).\nonumber\\
\eea

\subsection{Exchange energy}

In this section we obtain the exchange energy for a cylindrical as well as for a harmonic electron wire analytically, by integrating (\ref{exchE}). Specifically for cylindrical wire it turns out to be,
\bea
 E_x&=&-\frac{Nk_F}{\pi}\bigg(\frac{-1+2k_Fb\;K_1[2k_Fb]}{2(k_Fb)^2}+\pi K_0[2k_Fb]\nonumber\\
 & &\times L_{-1}[2k_Fb]+\pi K_1[2k_Fb]\;L_{0}[2k_Fb]\bigg) \label{Ex_Cyl}
\eea
where $K_n(x)$  is n$^{th}$ order modified Bessel function of
second kind, and $L_n(x)$ is modified Struve function\cite{Abramowitz72}. Similarly, the exchange energy can also be obtained for a harmonic wire of finite thickness given as
\bea
E_x&=&-\frac{Nk_F}{2\pi} \bigg(G_{2,3}^{2,2}\left(4 b^2 k_F^2|
\begin{array}{c}
 0,\frac{1}{2} \\
 0,0,-\frac{1}{2} \\
\end{array}
\right)\nonumber\\
& &-\frac{\ln\left(4 b^2 k_F^2\right)+e^{4 b^2 k_F^2}
   \Gamma \left(0,4 b^2 k_F^2\right)+\gamma }{4 b^2 k_F^2}\bigg),
\label{Ex_har}
\eea
where $G_{2,3}^{2,2}\left(4 b^2 k_F^2|
\begin{array}{c}
 0,\frac{1}{2} \\
 0,0,-\frac{1}{2} \\
\end{array}
\right)$ and $\Gamma \left(0,4 b^2 k_F^2\right)$  are the Meijer G function\cite{Bateman53} and the incomplete gamma function, respectively. For thin harmonic wires $b\ll a_B$ the exchange energy can be simplified to be,
\bea
 E_x&=&-\frac{Nk_F}{2\pi}  \bigg(
-1-\gamma-\ln(4)-2\ln(k_Fb)\nonumber\\
& &-2\; \psi^{(0)}(1/2)+2
\;\psi^{(0)}(3/2)\bigg),
\label{Ex_har1}
\eea
where $\psi^{(0)}(1/2)$ and  $\psi^{(0)}(3/2)$ are polygamma functions\cite{Abramowitz72}. We now use the simpler expansion of the polygamma function as
$2\;\psi^{(0)}(1/2)=-2\gamma-2\ln(4)$ and
$2\;\psi^{(0)}(3/2)=4-2\gamma-2\ln(4)$ in the above equation. The Eqs.(\ref{Ex_Cyl}) and (\ref{Ex_har}) can also be written for a polarized gas by defining $k_{F_{\uparrow(\downarrow)}}=k_F(1\pm p)$,
$N_{\uparrow(\downarrow)}=N(1\pm p)/2$ and $k_F=\pi/(2g_sr_s a_B)$. Explicitly for thin cylindrical wires $b\ll a_B$, the exchange energy per particle can be obtained by expanding Eq. (\ref{Ex_Cyl}) as
\bea
\label{excyl}
\epsilon_x&=&-\frac{1}{4g_sr_s}\bigg( (1+p)^2 \bigg[\frac{3}{2}-\gamma-\ln\bigg(\frac{\pi(1+p)}{2g_sr_s}\bigg)+\mathcal{L}\bigg]\nonumber\\
& &+
(1-p)^2\bigg[\frac{3}{2}-\gamma-\ln\bigg(\frac{\pi(1-p)}{2g_sr_s}\bigg)+\mathcal{L}\bigg]\bigg). \eea
Similarly for harmonic wires, Eq. (\ref{Ex_har1})  gives
\bea
\label{exhar}
\epsilon_x&=&-\frac{1}{4g_sr_s}\bigg( (1+p)^2\bigg[\frac{3}{2}-\frac{\gamma}{2}-\ln(2)-\ln\bigg(\frac{\pi(1+p) }{2g_sr_s} \bigg)\nonumber\\
& &+\mathcal{L} \bigg]+(1-p)^2
\bigg[\frac{3}{2}-\frac{\gamma}{2}-\ln(2)-\ln\bigg(\frac{\pi(1-p)}{2g_s
r_s}\bigg)\nonumber\\
& &+\mathcal{L} \bigg]\bigg), \eea
where $\mathcal{L}=\ln(a_B/b)$. It is noted that Eqs.(\ref{Ex_Cyl}) and (\ref{Ex_har}) are  new results and for special cases noted above they reduce to (\ref{excyl}) and (\ref{exhar}). It is worth mentioning that the logarithmic thickness of the wire is defined by $\mathcal{L}^{-1}$. For polarized ($g_s=1$ and $p=1$) and unpolarized fluids ($g_s=2$ and $p=0$), the exchange energy of a cylindrical wire is  obtained respectively to be

\be
\epsilon_x=-\frac{\ln(r_s)}{r_s}-\frac{1}{r_s}\bigg[ \frac{3}{2} -\gamma-\ln(\pi) + \mathcal{L}\bigg],
\ee
and
\be
\epsilon_x=-\frac{\ln(4r_s)}{4r_s}-\frac{1}{4r_s}\bigg[ \frac{3}{2} -\gamma-\ln(\pi) + \mathcal{L}\bigg].
\ee
These are in agreement with Fogler's results \cite{Fogler05a}.

\subsection{Correlation energy}

\begin{table*}[!t]
\footnotesize
\begin{center}
\caption{\footnotesize Parameters obtained in fitting correlation energy data with the formula given
in (\ref{appfunc1}) and (\ref{appfun2}) for various $r_s$.}\label{table1}
\begin{tabular}{ccccccccccc}
\hline
\hline
$r_s$&Function & $\alpha$ & $\beta$ & $\eta$ & $\chi^2$& AdjustedRSquared & AIC & BIC & RSquared &\\
\hline
0.1& $\epsilon_{c1}$ &-0.500 & -5.0&-9.15692 & -&- &-&-&-& Fig.2a \\
   & $\epsilon_{c2}$ &0.52184 & 14.9682&54.0315 & 0.997365&0.998632 & 291.265&298.579&0.998721&Fig.2a\\
   & $\epsilon_{c}$ &-0.000845452 & 0.319619&1.31095 &1& 0.999768& -279.783&-272.469&0.999784& Fig.2b\\
0.01& $\epsilon_{c}$ &-0.000378126 & -0.032961&-0.119755 &1 &0.999925 & -544.096&-536.781&0.99993&Fig.2c\\
0.001   & $\epsilon_{c}$ &-0.0000108161& -0.00244274&-0.0110844 &0.937348 &0.99788 &-333.214&-329.035&0.998183&Fig.2d\\
\hline
\end{tabular}
\end{center}
\end{table*}

\begin{figure}[!t]
\subfloat[{}]
{\includegraphics [scale=0.67]{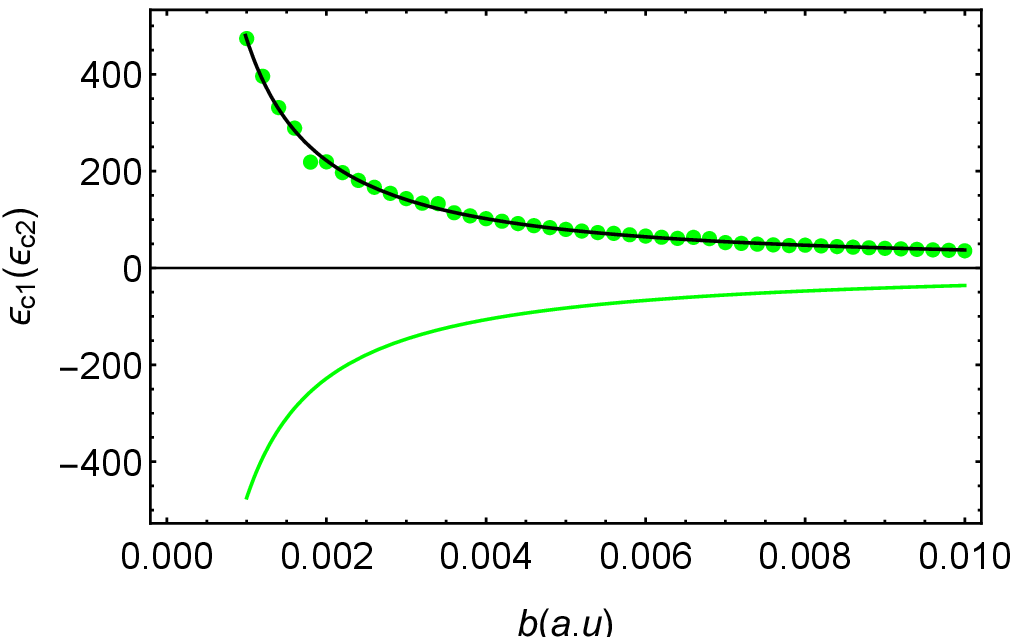}}\quad
\subfloat[{}]
{\includegraphics [scale=0.67]{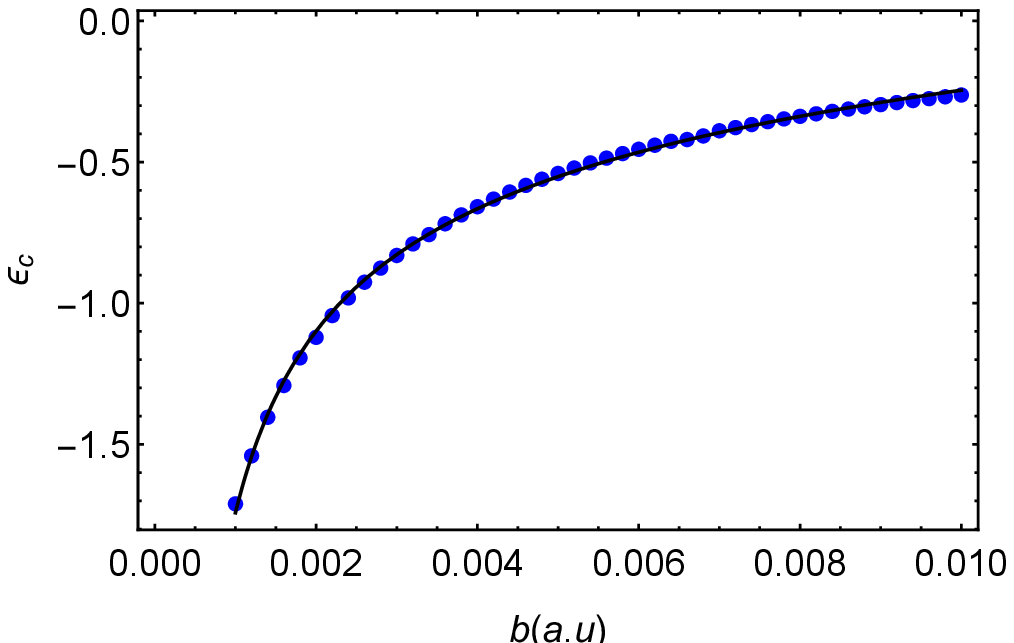}}\quad
\subfloat[{}]
{\includegraphics [scale=0.54]{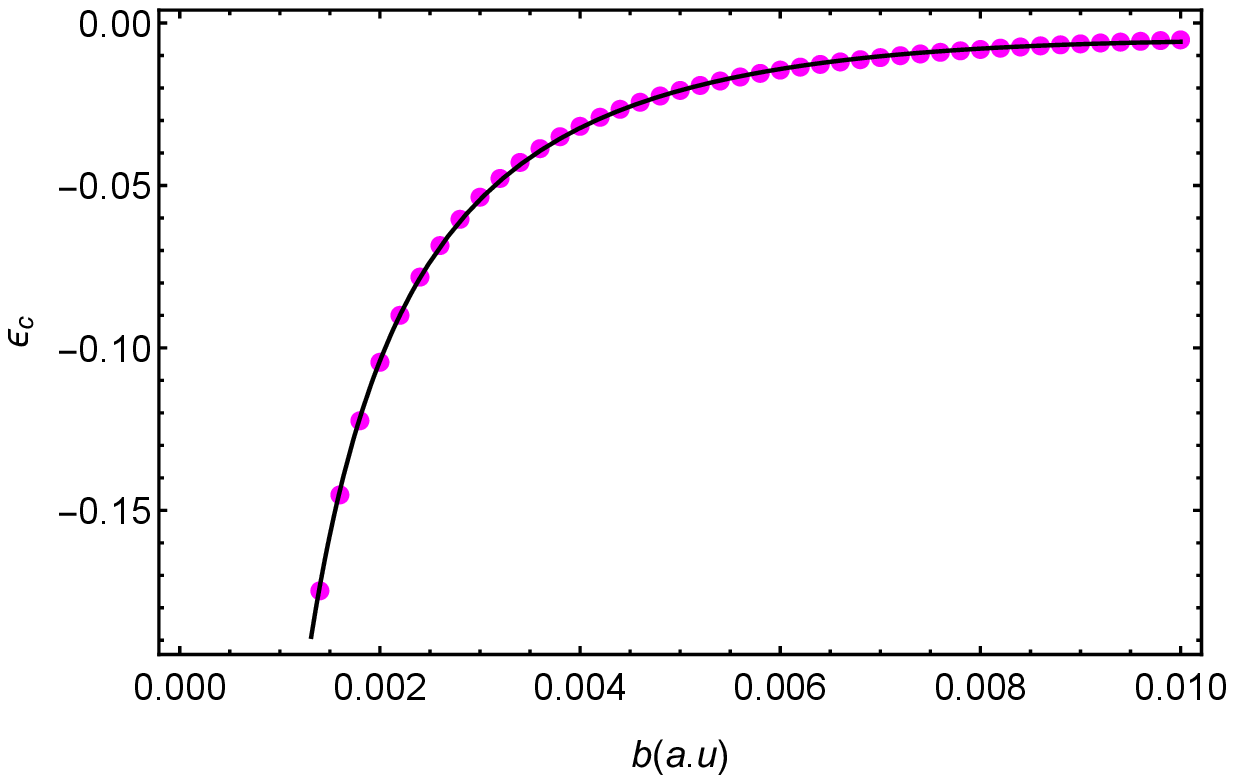}}\quad
\subfloat[{}]
{\includegraphics [scale=0.54]{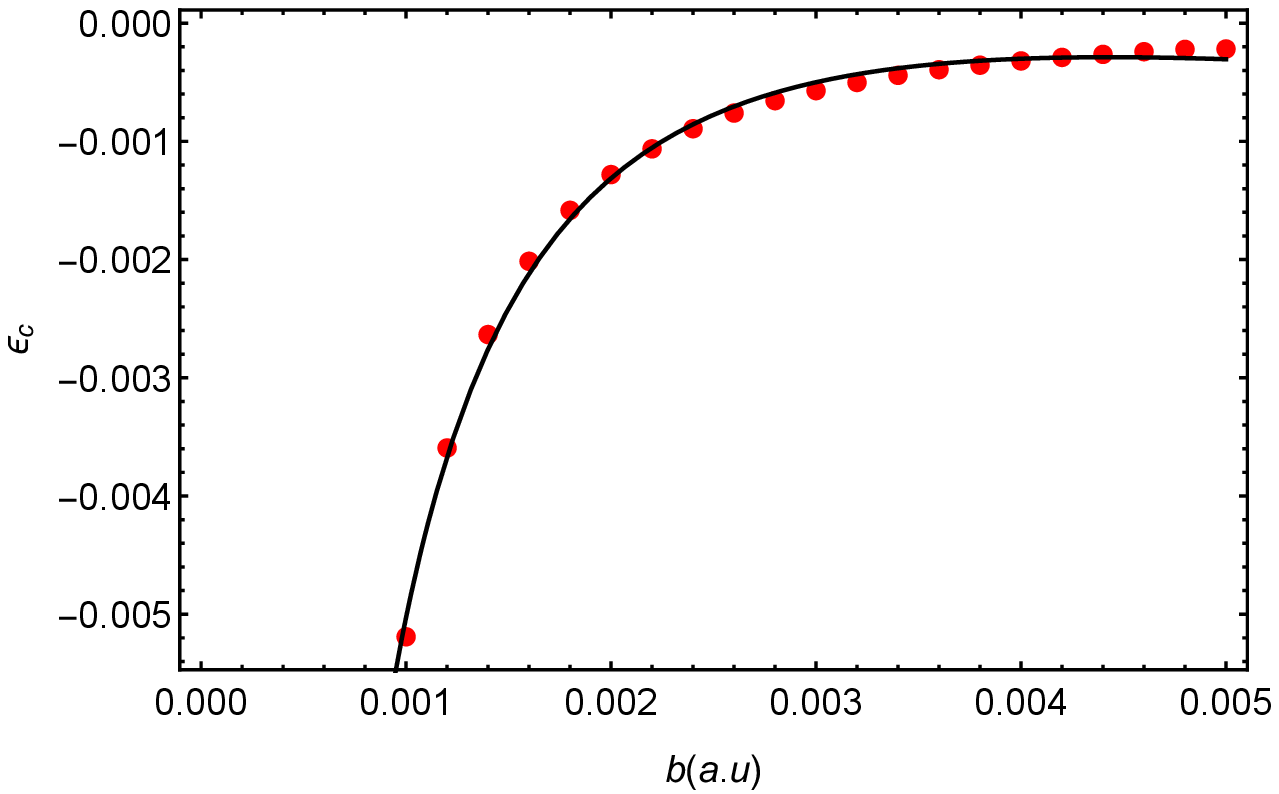}}
\caption{\footnotesize(color online) $(a)$ The parts of correlation energy $\epsilon_{c1}$ (lower curve) and $\epsilon_{c2}$ (upper curve) versus $b$ for $r_s=0.1$ where the analytical result $\epsilon_{c1}$ (green continuous line) are plotted together with the numerical $\epsilon_{c2}$ (green dots) and the fitted curve (black continuous line) for the same $r_s$. $(b)$ Total correlation energy $\epsilon_{c}$ for the same value of $r_s$ with the fitted curve (black continuous line) and the numerical results (blue dots). $(c)$  and $(d)$  for $r_s=0.01$ and $r_s=0.001$ respectively, but numerical results (dots) are shown by magenta and red colors.  }
\label{Ecorrs0.1}
\end{figure}

The integration over the coupling constant $\lambda$ is easily done in (\ref{corrE}) and the correlation energy  becomes
\bea
\label{corrEner}
E_c&=&\frac{n}{2}\sum_{q\neq0}\bigg(\frac{1}{n\pi}\int^{\infty}_0 \{v(q)\; \chi_0(q,i\omega)\nonumber\\
& &+\ln[1-v(q)\;\chi_0(q,i\omega)]\}d\omega \bigg).
\eea
The above equation can be written further as,
\bea
\epsilon_c=\epsilon_{c1}+\epsilon_{c2}
\eea
where
\bea
\epsilon_{c1}&=&-\frac{g_s}{2\pi} \int^{\infty}_0 v(q)\; S_0(q)\;dq\\
\label{ec4}\epsilon_{c2}&=&\frac{g_s}{2\pi}\int^{\infty}_0  \bigg(\frac{1}{n\pi} \int^{\infty}_0 ln\bigg\{ 1-v(q)\; \chi_0(q,i\omega) \bigg\}d\omega\ \bigg)dq.\nonumber\\
\eea
The first term $\epsilon_{c1}$ can be integrated analytically for the cylindrical potential,
\bea
\label{CorrEsmallb}
\epsilon_{c1}&=&-\frac{g_s^2 r_s a_B^2}{b^2\pi^2}+\frac{g_s a_B}{b\pi}K_1\bigg(\frac{b}{a_B}\frac{\pi}{g_s r_s}\bigg)+
\frac{a_B}{2 br_s}\nonumber\\
& \times&\bigg[ -g_s r_s+\pi (b/a_B) L_{-1}\bigg(\frac{b}{a_B}\frac{\pi}{g_s r_s}\bigg) K_0\bigg(\frac{b}{a_B}\frac{\pi}{g_s r_s}\bigg)\nonumber\\
& +&\pi b L_0 \bigg(\frac{b}{a_B}\frac{\pi}{g_s r_s}\bigg) K_1\bigg(\frac{b}{a_B}\frac{\pi}{g_s r_s}\bigg)\bigg].
\eea
The Eq. (\ref{CorrEsmallb}) is further simplified for an infinitely thin wire $b\rightarrow0$ for any finite $r_s$  as
\bea
\label{smallbEcorr}
\epsilon_{c1}=\frac{-g_s}{2(b/a_B)}+\frac{1}{2 r_s}\bigg[ \frac{3}{2}-\gamma+\ln\bigg(\frac{a_B}{b}\bigg)-\ln\bigg(\frac{\pi}{2g_sr_s} \bigg)\bigg].\nonumber\\
\eea

\begin{table*}[]
\footnotesize
\begin{center}
\caption{\label{LSCOTable1} \footnotesize Correlation energy per particle for fully polarized
fluids for various $r_s$ and $b$.}\label{table2}
\begin{tabular}{c|ccccccc}
\hline
\hline
&\multicolumn{7}{c}{correlation energy $\epsilon_c$ (mHartree) }\\
\cline{2-8}
$r_s$\;/\;b &0.001&0.01&0.1&0.2&0.3&0.4&0.5 \\
\hline
0.001&-5.1925&-0.051326&-0.00051321&-0.000128314&-0.000057025&-0.00032081&-0.000205298\\
0.01&-267.9063&-5.18143&-0.0512140&-0.01280232&-0.00568998&-0.00320059&-0.002048385\\
0.1&-1675.0533&-258.96748&-5.075932&-1.2577678&-0.5580828&-0.3137399&-0.20074031\\
0.2&-2250.7232&-568.01835&-19.866556&-4.9634643&-2.1935919&-1.2308436&-0.7868883\\
0.3&-2543.4151&-684.53832&-41.492422&-11.018708&-4.8636832&-2.7297174&-1.7380320\\
0.4&-2714.8034&-814.49113&-66.343762&-19.104517&-8.5835144&-4.7704732&-3.0480646\\
0.5&-2815.2034&-895.4&-93.330872&-28.570889&-13.510079&-7.3493536&-4.6830739\\
\hline
\end{tabular}
\end{center}
\end{table*}

For a given $r_s$ the above equation has a functional dependence on $b$ (in atomic unit) as
\bea
\label{appfunc1}
\epsilon_{c1} (b,r_s)= \frac{\alpha(r_s)}{b} + \beta(r_s)\; \ln(b) + \eta(r_s),
\eea
where $\alpha(r_s)$, $\beta(r_s)$ and   $\eta(r_s)$ can be read-off (\ref{smallbEcorr}). Eq.(\ref{ec4}) cannot be integrated analytically, therefore we solve it numerically. Anticipating that the correlation energy $\epsilon_c$  for $b\to 0$ and $r_s\to 0$ turns out to be a constant, $\epsilon_{c2}$ may also be represented by  (\ref{appfunc1}) with the same coefficient $\alpha(r_s)$, $\beta(r_s)$ but with a differing sign and a different constant  $\eta(r_s)$. Therefore we represent $\epsilon_{c2}$ by,
\bea
\label{appfun2}
\epsilon_{c2} (b,r_s)= \frac{\alpha(r_s)}{b} + \beta(r_s)\; \ln(b) + \eta(r_s),
\eea
and fit it to the numerical result. The coefficients $\alpha$, $\beta$, $\eta$ for  $\epsilon_{c2}$ and $\epsilon_{c}$ are given in Table (\ref{table1}), for $r_s$=0.1, 0.01 and 0.001. Note that the coefficients for  $\epsilon_{c1}$ are analytically known. Also the same formula as for $\epsilon_{c2}$ is assumed for $\epsilon_{c}$. To estimates the accuracy of the fit with the numerical calculation, we have provided the statistical analysis with different methods:  $\chi^2$, $R^2$ adjusted for the number of model parameters (AdjustedRSquared), Akaike information criterion (AIC), Bayesian information criterion (BIC)  and coefficient of determination $R^2$ (RSquared). The fitted parameters by the statistical analysis in Table(\ref{table1}) reflect the quality of the function $\epsilon_{c2} (b,r_s)$ and $\epsilon_{c} (b,r_s)$.
%

\begin{figure}[!b]
\centering
\includegraphics[scale=0.65]{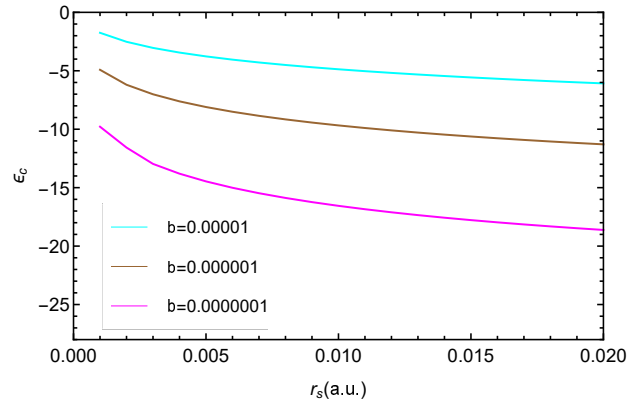}
\caption{\footnotesize(Color online) The correlation energy per particle is plotted versus $r_s$ for different thickness $b$ of cylindrical wire.}\label{Ecorvsrs}
\end{figure}


\begin{figure}[!b]
\centering
\includegraphics[scale=0.65]{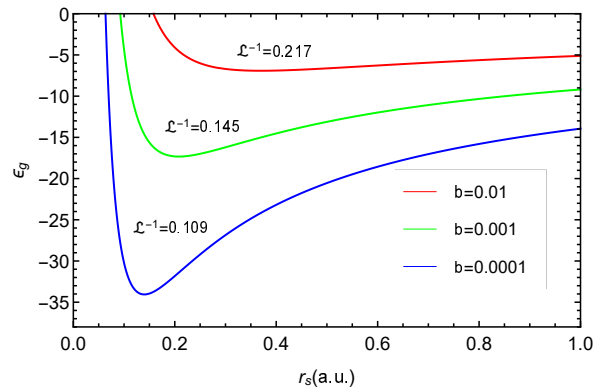}
\caption{\footnotesize(Color online) Ground state energy $\epsilon_g$ is plotted as a function of $r_s\leq1$, for values of wire widths  $b$.}\label{GRSvsrs}
\end{figure}


The correlation energies per particle $\epsilon_{c1}$ (lower curve) and $\epsilon_{c2}$ (upper curve) are plotted in Fig. \ref{Ecorrs0.1}a, as obtained analytically and numerically, and are shown as green continuous curve and green dots, respectively. The fitted $\epsilon_{c2}$ from representation  (\ref{appfun2}) is shown by the black continuous line. It is clearly seen that there is a perfect fit of $\epsilon_{c2}$, as also inferred above from the statistical analysis. It is seen that there is no cancellation between the two curves for  $r_s=0.1$. The resulting sum is plotted for the same $r_s$ in Fig.\ref{Ecorrs0.1}b. Total  correlation energy for $r_s=0.01$ and $r_s=0.001$ are also plotted in Fig. \ref{Ecorrs0.1}c and Fig.\ref{Ecorrs0.1}d respectively. These figures show that there is no indication that the correlation energy approaching a constant value for very small $r_s$ for an infinitely thin wire. Rather it diverges contrarily to the result obtained by Loos\cite{Loos13}as $b$ become vary small.

For a pragmatic width of the wire, the correlation energy for a polarized fluid is reported in Table \ref{table1}. The correlation energy at high densities  $r_s\leq0.1$ and $b=0.1$, is in agreement with the quantum Monte Carlo simulation \cite{Casula06,Malatesta99} for polarized fluids.

To check the consistency of our result of the correlation energy for $b\to0$ and $r_s<1$, we plot it in Fig. \ref{Ecorvsrs}  for small values of $b$ shown therein as a function of $r_s$. It is seen from Fig. \ref{Ecorvsrs}  that as $b$ decreases, the correlation energy increases, which is consistent with our previous results given in Figs. \ref{Ecorrs0.1}b , 2c  and Fig. \ref{Ecorrs0.1}d.

In Fig. \ref{GRSvsrs} we also plot the total ground-state energy with different wire thicknesses as a function of $r_s$. It is noted that as $b$ decreases, the ground state energy increases. There are no QMC data available to compare the ground-state energy for these $r_s\ll 1$, for an infinitely thin wire. It is pointed out that our calculation is suited for long-range interactions whereas the Fogler calculation deals with the short-range interaction.

\section{Summary}
\label{Result_Discussion}

In this paper we have calculated the dependence of the ground-state structure factor and the correlation energy on the thickness of an electron wire as a function of $r_s$. The structure factor is calculated in the single-loop approximation of RPA. The electron-electron interactions are modeled by a cylindrical and a harmonic potential. We find an agreement with the result obtained by Fogler\cite{Fogler05} by a variational calculation.  The structure factor has also been compared for $b=0.025$ and $r_s=0.1$ with the QMC data\cite{vinod17}. For first-order corrections in the interaction, the RPA results and the QMC data match perfectly, indicating that for small thickness and for high densities, the electron gas behaves as a gas of non-interacting particles but highly correlated which is clear from the correlation energy calculations. In this sense Fogler\cite{Fogler05,Fogler05a} calls it a Coulomb Tonks gas.

We have also obtained the exchange energy for both cylindrical and  harmonic electron wires analytically. These expressions are new. In the small-thickness limit the expressions simplify considerably and are more or less the same for both wires. This has been also worked out for polarized gases, from which the paramagnetic and ferromagnetic phases can easily be obtained. It is also noted that the exchange energy for a fully polarized gas agrees with Fogler\cite{Fogler05a}.


%

In the present paper
the total correlation energy in RPA are found to be fitted by
\bea
\epsilon_c (b,r_s)= \frac{\alpha(r_s)}{b} + \beta(r_s)\; \ln(b) + \eta(r_s)
\label{final}
\eea
with the parameters given explicitly.
This correlation energy is the sum of two terms
which only partially cancel.
The first term is calculated analytically exactly
by the expression (\ref{final})
where the values of $\alpha$, $\beta$ and $\eta$ are precisely known.
The second term has been calculated numerically. It perfectly fits with the expression of (\ref{final}) but with different parameters.

This findings clearly indicate that the correlation energy is diverging in the limit of $b\to 0$ and $r_s\to0$, in contrary to the conventional perturbation theory result \cite{Loos13}. Further, the correlation energy as a function of $r_s$ for various $b$, again points out that the correlation energy increases as $b$ decreases for $r_s\to 0$. The Coulomb correlations are enhanced, and the interacting electron gas behaves structure-less in the ultrathin and high-density domain of $\mathcal{L}^{-1}\ll r_s\ll 1$ like a strongly-interacting electron gas named Coulomb-Tonks gas (CTG)\cite{Fogler05,Fogler05a}. Further, we find that the correlation energy does not approach a constant value for an infinitely thin wire and  $r_s\to 0$ within the RPA.

\begin{acknowledgments}
The authors(VA and KNP) acknowledge the financial support by National Academy of Sciences of India.
\end{acknowledgments}

\end{document}